# ASTEROID PERTURBATIONS AND MASS DETERMINATION FOR THE ASTROD SPACE MISSION

Chien-Jen Tang[1] and Wei-Tou Ni[1]

[1]*Center for Gravitation and Cosmology,
Department of Physics, National Tsing Hua University,
Hsinchu, Taiwan 30055, ROC*

**ABSTRACT**

Due to the high-precision nature of the ASTROD (Astrodynamical Space Test of Relativity using Optical Devices) mission concept, the asteroid perturbations on the ASTROD spacecraft is crucial. These perturbations need to be modelled and determined together with relativistic effects and other solar-system parameters. In a previous investigation (Su *et al*., *Planetary and Space Science*, **47**, 339-43[1999]), we used the mass estimation of Ceres, Pallas and Vesta in the literature to calculate their perturbations on the ASTROD spacecraft. Recently, we established an ephemeris framework (CGC 1) including the 3 big asteroids and used this ephemeris framework to simulate the determination of their masses together with other solar-system parameters and relativistic-gravity parameters. In this paper, we extend the CGC 1 to CGC 2 ephemeris framework to include 492 asteroids (with diameter > 65 km) . We then use CGC 2 to simulate the determination of ten parameters --- the masses of Ceres, Pallas and Vesta, the six average densities for the other 489 asteroids, classified into C, S, M, E, G and U types, and the rate of change of the gravitational constant G. The fractional mass uncertainties for Ceres, Pallas and Vesta are about $10^{-3}$-$10^{-4}$ ; the fractional density uncertainties for six types of asteroids are about $10^{-2}$-$10^{-3}$ ; the uncertainty for the determination of $\dot{G}/G$ is about $10^{-14}$-$10^{-15}$/yr. This mass and density determination will be useful for the determination and understanding of the structure and origin of asteroids.

**INTRODUCTION**

We have been studying the Astrodynamical Space Test of Relativity using Optical Devices (ASTROD) mission concept since 1993 (Ni, 1993; Ni, Wu and Shy, 1996; Ni *et al*., 1997). The objectives of ASTROD mission are threefold. The first objective is to discover and explore fundamental physical laws governing matter, space and time via testing relativistic gravity with 3-6 orders of magnitude improvement. The second objective of the ASTROD mission is the high-precision measurement of the solar-system parameters. The third objective is to detect and observe gravitational waves from massive black holes and galactic binary stars in the frequency range 50 µHz to 5 mHz. Background gravitational-waves will also be explored. In January, 2000, we formed an international team and submitted the ASTROD proposal to ESA (Bec-Borsenberger *et al*., 2000).

A desirable implementation is to have two spacecraft in separate solar orbits each carrying a payload of a proof mass, two telescopes, two 1-2 W lasers, a clock and a drag-free system, together with an Earth reference system. The Earth reference system could be ground stations, Earth satellites and/or spacecraft near Earth-Sun Lagrange points. For technological development, please see Bec-Borsenberger *et al*. (2000) and references therein.

Due to the ultrahigh precision of ranging in the ASTROD mission, the asteroid perturbations of the ASTROD spacecraft is important both in themselves (in the determination of asteroid parameters) and in the determination of relativistic parameters and other solar-system parameters. In a previous investigation (Ni, 1997), we proposed that the temporal variation in the gravitational constant can be measured to $10^{-13}$/yr or better in fraction in the ASTROD mission. This depends largely on the separability of the influence of asteroidal perturbations. In a subsequent paper (Su *et al*., 1999), we used the mass estimation of Ceres, Pallas and Vesta in the literature to calculate the perturbations of the spacecraft and propose to determine the masses of asteroids through their perturbations on the ASTROD spacecraft.

To go further, we started orbit simulation and parameter determination (Chiou and Ni, 2000a, b). We worked out a post-Newtonian ephemeris of the Sun, the major planets and 3 biggest asteroids including the solar quadrupole moment. We term this working ephemeris CGC 1 (CGC: Center for Gravitation and Cosmology). Using this ephemeris as a deterministic model and adding stochastic terms to simulate noise, we generate simulated ranging data and use Kalman filtering to determine the accuracies of fitted relativistic and solar-system parameters after 1050 days of the mission as follows:

$\gamma$, $4.6 \times 10^{-7}$;  $M_{ceres}$, $6.4 \times 10^{-5} M_{ceres}$;
$\beta$, $4.0 \times 10^{-7}$;  $M_{pallas}$, $7.6 \times 10^{-4} M_{pallas}$;
$J_2$, $1.2 \times 10^{-8}$;  $M_{vesta}$, $8.1 \times 10^{-5} M_{vesta}$.

When we add the parameters $\dot{G}/G$ and $a_a$ in the determination where $a_a$ is the anomalous acceleration toward the Sun of Anderson *et al* (1998), the accuracies of their values after fitting are $9.5 \times 10^{-15}$/yr and $2.0 \times 10^{-16}$ m/s$^2$.

For a better evaluation of the accuracy of $\dot{G}/G$, we need also to monitor the masses of other asteroids. For this, we consider all presently known 492 asteroids with diameter greater than 65 km, obtain an improved ephemeris framework --- CGC 2, and calculate the perturbations due to these 492 asteroids on the ASTROD spacecraft. We then add stochastic terms to simulate ranging data and to determine the expected accuracy of the parameter $\dot{G}/G$ and the nine fitted parameters of these 492 asteroids. These nine parameters are the masses of Ceres, Pallas and Vesta and six average densities for the other 489 asteroids, classified into C, S, M, E, G and U types according to apparent surface composition as in Table 1. The classification in Table 1 is taken from Bowell (1999) with references to Tedesco (1989) and Tholen (1984). We take five classes — C, S, M, E and G. All other/unknown classes are grouped into the U class. A file is kept in CGC for reference. For the densities of C, S, M classes, we use the same values as in JPL DE 405 (Standish, 1998). For the density of E class, we adopt the average value of Wasson (1974). Since the G class can be consider as a subclass of C class, for its density we use the C class value. The density for the class U of unknown/other composition is estimated using the number weighted mean of the other five classes.

**Table 1. Classification of 489 Asteroids into 6 Types (Bowell, 1999) and the Accuracy of Determination of their Average Densities in this ASTROD Simulation**

| Type | Symbol | Number | Input Density | Nominal Total Mass | Fractional Accuracy in the ASTROD Simulation |
|---|---|---|---|---|---|
| Carbonaceous Chondritic | C | 144 | 1.8 g/cm$^3$ | 6.54×10$^{-14}$ | $8.18 \times 10^{-4}$ |
| Silicaceous/Stone-iron | S | 52 | 2.4 g/cm$^3$ | 2.28×10$^{-14}$ | $8.42 \times 10^{-4}$ |
| Metallic (nickel-iron) | M | 14 | 5.0 g/cm$^3$ | 1.31×10$^{-14}$ | $1.33 \times 10^{-3}$ |
| Enstatite Achondritic | E | 2 | 3.65 g/cm$^3$ | 1.84×10$^{-16}$ | $5.53 \times 10^{-3}$ |
| Extreme UV Feature* | G | 5 | 1.8 g/cm$^3$ | 3.07×10$^{-15}$ | $4.29 \times 10^{-3}$ |
| Other/Unknown | U | 272 | 2.167 g/cm$^3$ | 6.59×10$^{-14}$ | $7.07 \times 10^{-4}$ |

*The G class can be considered as a subclass of C class. For the present simulation, we take this subclass separately to see how the subclass density can be determined.

The accuracies of mass determination for Ceres, Pallas and Vesta are $7.77 \times 10^{-5}$ $M_{ceres}$, $6.25 \times 10^{-4}$ $M_{pallas}$ and $9.40 \times 10^{-5}$ $M_{vesta}$. The accuracy of the determination of $\dot{G}/G$ is $2.56 \times 10^{-15}$/yr. These figures are comparable to the simulation of our previous work (Chiou and Ni, 2000b). The differences of two simulations are due to differences in the set of the parameters evaluated and the duration of mission. In the present simulation, we have not include relativistic parameters except $\dot{G}/G$ although more asteroids are considered, and the duration of mission is 1200 days instead of 1050 days.

In Section 2, we discuss the motion of asteroids, establish CGC 2 ephemeris and compare CGC 2 ephemeris with DE 403 (Standish et al., 1995) and DE 405 ephemerides. In Section 3, we calculate the perturbations of 492 asteroids on the ASTROD spacecraft. In Section 4, we generate the simulated ranging data and use Kalman filtering to calculate the accuracies of 9 asteroid parameters and $\dot{G}/G$ to be fitted. At the end, we conclude with a discussion.

## CGC 2 EPHEMERIS

In ASTROD orbit simulation and determination of relativistic and solar-system parameters, we need a working ephemeris that we can add parameters and terms to their generation. For this purpose, we have built and used CGC 1 ephemeris in Chiou and Ni (2000a, b). For CGC 1 ephemeris, we used the following barycentric metric with solar quadrupole moment

$$ds^2 = [1 - 2\sum_i \frac{m_i}{r_i} + 2\beta(\sum_i \frac{m_i}{r_i})^2 + (4\beta - 2)\sum_i \frac{m_i}{r_i}\sum_{j \neq i} \frac{m_i}{r_{ij}} - c^{-2}\sum_i \frac{m_i}{r_i}(2(\gamma+1)\dot{x}_i^2 - r_i \cdot \ddot{x}_i - \frac{1}{r_i^2}(r_i \cdot \dot{x}_i)^2)$$
$$+ \frac{m_1 R_1^2}{r_1^3} J_2 (3(\frac{r_1 \cdot \hat{z}}{r_1})^2 - 1)]c^2 dt^2 + 2c^{-1}\sum_i \frac{m_i}{r_i}((2\gamma+2)\dot{x}_i) \cdot dx c dt - [1 + 2\gamma \sum_i \frac{m_i}{r_i}](dx)^2, \qquad (1)$$

where $r_i = x - x_i$, $r_{ij} = x_i - x_j$, $m_i = GM_i/c^2$, and $M_i$'s the masses of the bodies with $M_1$ the solar mass (Brumberg, 1991). $J_2$ is the quadrupole moment parameter of the Sun. $\hat{z}$ is the unit vector normal to the elliptic plane. The associated equations of motion of N-mass problem is derived from the geodesic variational principle of this metric and is used to build our computer-integrated ephemeris (with $\gamma = \beta = 1$, $J_2 = 2 \times 10^{-7}$) for nine-planets, the Moon and the Sun. The positions and velocities at the epoch 2005.6.10 0:00 are taken from the DE403 ephemeris. The evolution is solved by using the 4$^{th}$-order Runge-Kutta method with the stepsize h =0.01 day. In Chiou and Ni (2000b), the 11 body evolution is extended to 14 body evolution to include the 3 big asteroids — Ceres, Pallas and Vesta (CGC 1 ephemeris).

To improve CGC 1 ephemeris, and to do a better simulation for the accuracy in the determination of $\dot{G}/G$ and asteroid masses and densities, we include additional 489 asteroids with diameter larger than 65 km in the calculation of the perturbation of orbits of nine planets, the sun, the moon and the 3 big asteroids. To simplify calculation, the *heliocentric* orbits of these 489 asteroids are determined by the Kepler elements listed in Bowell (1999). At each calculation step, the Newtonian perturbation forces are calculated and added to the equations of motion of the 14 celestial bodies. For more accurate calculation of the positions and velocities of the Earth and the Moon, the quadrupole moment effect of Earth is added to the equations of motion of the Earth and the Moon. This is the CGC 2 ephemeris framework. We use the Runge-Kutta 4$^{th}$ order algorithm to solve the differential equations of essential celestial bodies with the stepsize 0.01 days. The initial time is 0:00 2005/06/10 (JD 2453531.5) with the initial positions and velocities of 11 celestial bodies taken from JPL DE405 ephemeris; those of the three big asteroids are calculated from MPO98 (1997). Figure 1 shows a comparison of CGC 2 and DE 403 with DE 405 for the range, latitude and longitude of Mercury and Mars in the Earth-Moon mass-center equatorial coordinate frame for 1200 days, after the initial time. For CGC 2, the deviations from DE 405 are below 0.5 km in range, 0.3 mas in latitude and 1.2 mas in longitude.

## PERTURBATIONS ON THE ASTROD SPACECRAFT

In previous works, we have used DE403 ephemeris and CGC 1 ephemeris to evaluate the gravitational effect on the spacecraft orbits from three largest asteroids---Ceres, Pallas and Vesta (Su *et al*., 2000; Chiou and Ni, 2000b). Here we use the CGC 2 ephemeris to evaluate the perturbations in heliocentric coordinates. In calculating the perturbed orbit of the spacecraft, we use Runge-Kutta method of 4$^{th}$ order to solve the equations of motion for the two spacecraft with the asteroid perturbation forces added. We use the stepsize 0.01 days and the initial time 0:00 2005/06/10 (JD 2453531.5). The differences of asteroid-perturbed orbits and unperturbed orbits in heliocentric coordinates for the inner spacecraft and outer spacecraft are shown in Fig. 2. The perturbations due to Ceres agree with Su et al. (2000). However, the Fig.2(b) of this reference were taken from a wrong file (Su et al., private communication); When compared with the correct file, there is also an agreement.

## SIMULATION

We use the following stochastic model to simulate the ranging time between ASTROD spacecraft and Earth. We consider two kinds of noises. The first kind is the imprecision of timing of optical ranging devices. It will influence ranging time directly. This part is treated as a Gaussian random noise with zero mean and with magnitude $5 \times 10^{-11}$ sec. The second kind is unknown accelerations due to imperfection of the spacecraft drag free system. The magnitude of unknown acceleration is treated as a Gaussian random noise with zero mean and with half width $10^{-15}$ m/s$^2$. We change the direction of unknown acceleration randomly every four hours. (Chiou and Ni, 2000 a ,b)

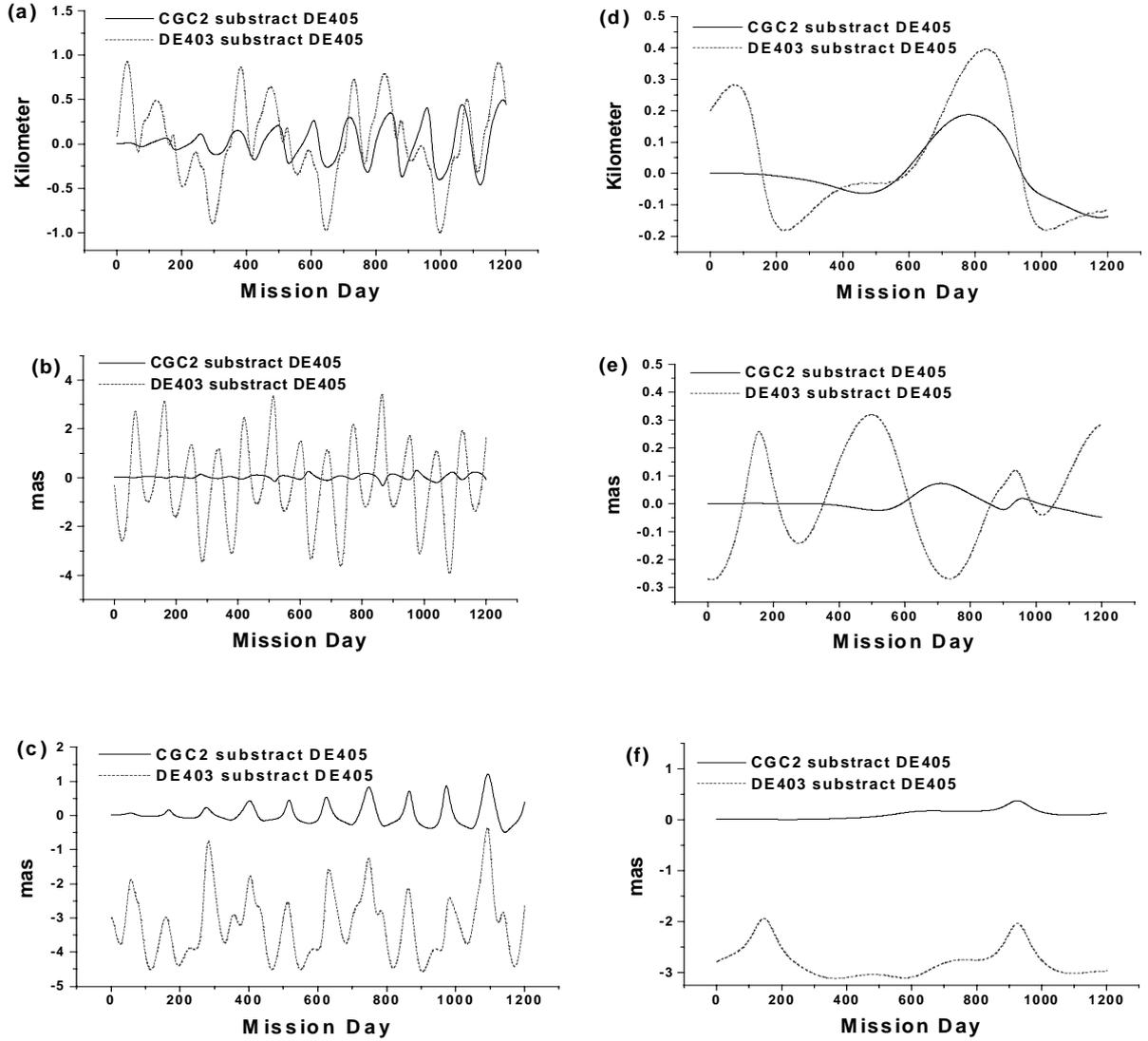

Fig. 1. Comparison of CGC 2 and DE 403 with DE 405 for the deviations of (a) range, (b) latitude and (c) longitude of Mercury, and the deviations of (d) range, (e) latitude and (f) longitude of Mars in the Earth-Moon mass-center equatorial coordinate frame for 1200 days. The initial time is 0:00 2005/06/10 (JD2453531.5).

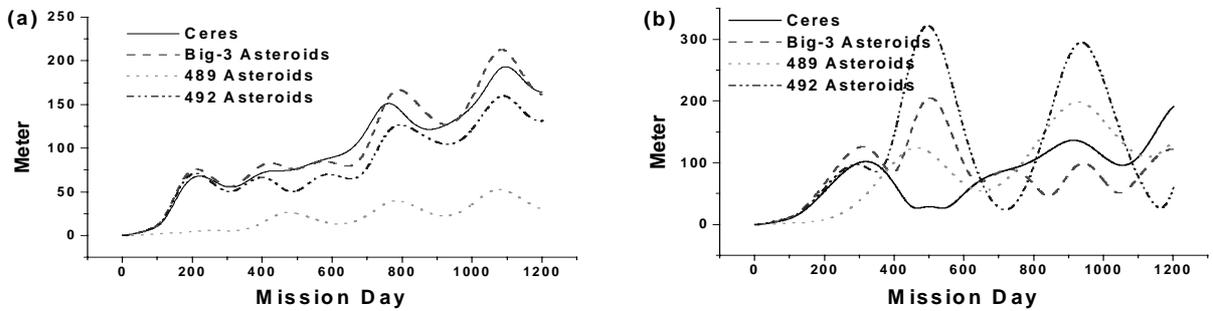

Fig. 2. Perturbations on (a) inner spacecraft and (b) outer spacecraft due to Ceres, the big 3 asteroids (Ceres, Pallas and Vesta), the 489 asteroids and the 492 asteroids (big 3 + 489 asteroids) in heliocentric coordinates.

In our simulation process, we first deal with acceleration noise and integrate it with equations of motion to obtain positions. Then Gaussian random timing noise are added to ranging time. The resulting simulation data are similar to Chiou and Ni (2000 a,b).

We adopt Kalman sequential filtering method for parameter fitting in the ASTROD mission. When we get a new observation $z = Ax + v$ where A is observation partial, $v$ is observation error, we can update our a priori estimate $\tilde{x}$ and a priori covariant matrix $\tilde{P}$ by Kalman filtering method. The diagonal term of $P$, $P(i,i)$, is the error square of the parameter $x_i$. We can compute the correlation of $x_i$ and $x_j$ from the off-diagonal term of $P(i,j)$.

In our work, the estimated parameter $x$ is a 10×1 matrix consisting of the masses of Ceres, Pallas and Vesta, the densities of C, S, M, E, G and U classes, and $\dot{G}/G$:

$$x = (M_{Ceres}, M_{Pallas}, M_{Vesta}, D_C, D_S, D_M, D_E, D_G, D_U, \dot{G}/G)^T,$$

where the superscript $T$ means taking the transverse. $z = R(x,t) + v(t)$ is a 1×1 observing matrix, where $R(x,t)$ is the ranging time computed by deterministic model, and $v(t)$ is the timing noise. But $R(x,t)$ is not linear function of $x$. So it must be linearized before we can use Kalman filtering method to fit $x$. Expand $R(x,t)$ to first order in $x_0$ by Taylor series,

$$z = z_0 + \nabla R(x_0,t) \cdot (x - x_0) + O((x - x_0)^2) + \cdots + v(t).$$

If we redefine $z \equiv z - z_0$, $A \equiv \nabla R(x_0,t)$, $x \equiv x - x_0$ and drop the higher order term (The linearization error is negligible.), then the equation can be rewritten as $z = Ax + v$. Using the sequential Kalman filter data processing (Bierman, 1977), we obtain the estimated uncertainties of these ten parameters to be determined from the mission as functions of epoch. Each day at 0:00, one ranging data point for the inner spacecraft and one ranging data point for the outer spacecraft are fitted. The initial uncertainties of the ten parameters are taken as following:

$\sigma_{M\,Ceres} = 0.5$, $\sigma_{M\,Pallas} = 0.5$, $\sigma_{M\,Vesta} = 0.5$, $\sigma_{D\,classC} = 0.2$, $\sigma_{D\,classS} = 0.2$,

$\sigma_{D\,classM} = 0.2$, $\sigma_{D\,classE} = 0.2$, $\sigma_{D\,classG} = 0.2$, $\sigma_{D\,classU} = 0.2$, $\sigma_{\dot{G}/G} = 1 \times 10^{-11}/yr$.

The uncertainties as functions of epoch are shown in Fig.3.

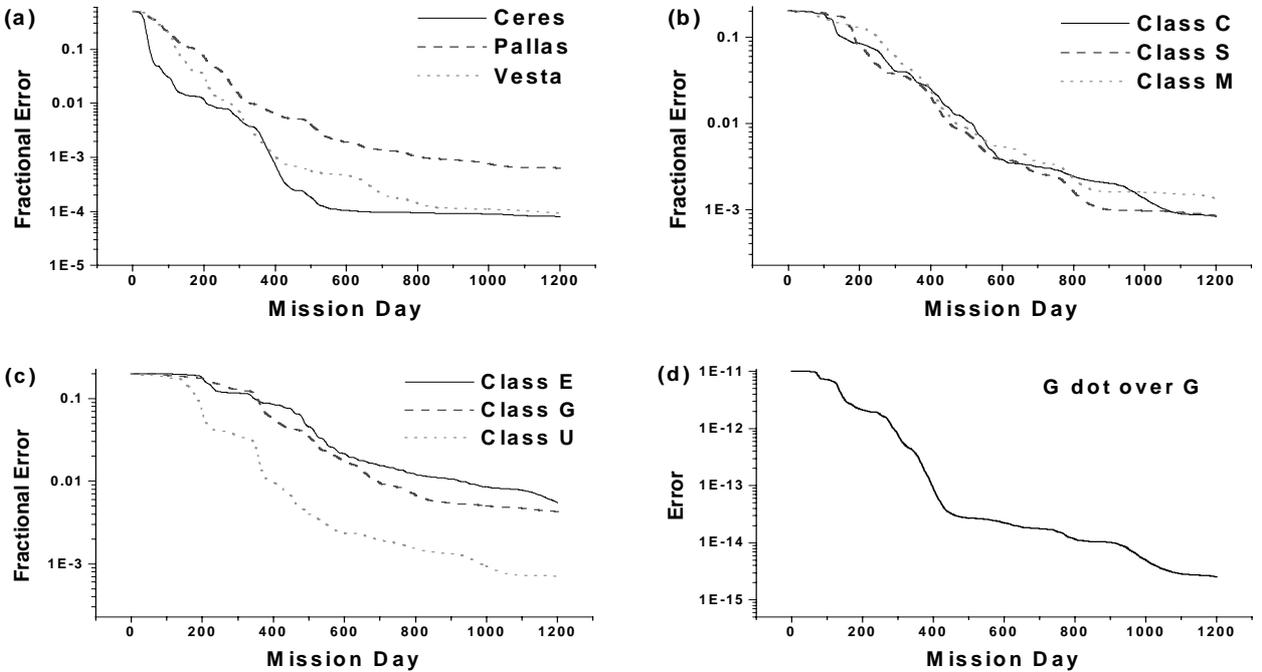

Fig. 3. The fractional uncertainties as functions of mission epochs for (a) Ceres, Pallas, Vesta masses, (b) class C, class S, class M densities, and (c) class E, class G, class U densities. (d) shows the uncertainty as a function of epoch in $\dot{G}/G$.

The final uncertainties after 1200 days of mission are:

$\sigma_{M\ Ceres} = 7.77 \times 10^{-5}$, $\sigma_{M\ Pallas} = 6.25 \times 10^{-4}$, $\sigma_{M\ Vesta} = 9.40 \times 10^{-5}$, $\sigma_{D\ classC} = 8.18 \times 10^{-4}$, $\sigma_{D\ classS} = 8.42 \times 10^{-4}$, $\sigma_{D\ classM} = 1.33 \times 10^{-3}$, $\sigma_{D\ classE} = 5.53 \times 10^{-3}$, $\sigma_{D\ classG} = 4.29 \times 10^{-3}$, $\sigma_{D\ classU} = 7.07 \times 10^{-4}$, $\sigma_{\dot{G}/G} = 2.56 \times 10^{-15}$ / yr.

**DISSCUSSION**

The simulation here confirms our former studies on the accuracies achievable in the determination of $\dot{G}/G$, and the masses of Ceres, Pallas and Vesta. The fractional density uncertainties for six types of asteroids are about $10^{-2}$-$10^{-3}$. This will be useful for modelling the structure and origin of asteroids.